\documentclass[pdftex,twocolumn,epjc3]{svjour3}          

\RequirePackage[T1]{fontenc}

\smartqed  
\RequirePackage{graphicx}
\RequirePackage{mathptmx}      
\RequirePackage{flushend}
\RequirePackage[numbers,sort&compress]{natbib}
\RequirePackage[colorlinks,citecolor=blue,urlcolor=blue,linkcolor=blue]{hyperref}
\RequirePackage{epstopdf}
\journalname{Eur. Phys. J. C}
\RequirePackage{amsmath,amssymb}
\begin{document}

\title{Fat brane, dark matter and localized kinetic terms in six dimensions}


\author{Ricardo G. Landim\thanksref{e1,addr1}
}

\thankstext{e1}{ricardo.landim@tum.de}

\institute{Physik Department T70, James-Franck-Stra$\beta$e 1,
Technische Universit\"at M\"unchen, 85748 Garching, Germany\label{addr1}
}

\date{Received: date / Accepted: date}

\maketitle

\begin{abstract}
Extra dimensions (ED) have been used as attempts to explain several phenomena in particle physics over the years. In this paper we investigate the role of an abelian gauge field as mediator of the interaction between dark matter (DM) and  Standard Model (SM) particles, in a model with two flat and transverse ED compactified on the chiral square. DM is confined in a thin brane, localized at the origin of the chiral square, while the SM is localized in a finite width brane, lying in the opposite corner of the square. A brane-localized kinetic term   is present in the DM brane, while in the fat brane  it is not allowed.   In this model the kinetic mixing is not required because we assume that the SM 
particles couple to the mediator through  their $B-L$ charges, while DM couples to it  via a dark charge. Assuming a complex scalar field as DM candidate it is possible to obtain the observed DM relic abundance and avoid direct detection constraints for some parameter choices.   \end{abstract}

\section{\label{sec:intro}Introduction} 
Weakly Interacting Massive Particles (WIMPs) have been the most well-known 
dark matter (DM) candidates \cite{Arcadi:2017kky} for decades, but the absence of any trace encourage us to look for different scenarios, both   experimentally and theoretically. One promising way to chase DM would be if it interacts with the Standard Model (SM) particles through a new mediator.  A relatively recent and very explored idea is the possible interaction between DM and SM via a  new dark $U(1)_D$ gauge field, arising in turn from a kinetic mixing term between this new vector mediator (called dark photon, DP) and the
hypercharge $U(1)_Y$ field \cite{Holdom:1985ag,Holdom:1986eq,Dienes:1996zr,DelAguila:1993px,Babu:1996vt,Rizzo:1998ut,Feldman:2006wd,Feldman:2007wj,Pospelov:2007mp,Pospelov:2008zw,Davoudiasl:2012qa,Davoudiasl:2012ag,Essig:2013lka,Izaguirre:2015yja,Curtin:2014cca,Davoudiasl:2013aya,Kim:2019oyh,Hook:2010tw}.

Among other theoretical alternatives, extra dimensions (ED) have been considered over the decades as tools to address a wide range of issues in particle physics, such as   the  hierarchy
 \cite{Antoniadis:1990ew,Dienes:1998vh,Antoniadis:1998ig,ArkaniHamed:1998rs,Randall:1999ee,Arkani-Hamed:2016rle,Arun:2017zap,Arun:2018yhg} and flavor problems \cite{Agashe:2004cp,Huber:2003tu,Fitzpatrick:2007sa}. Models employing two ED, for example, may provide explanations for proton stability \cite{Appelquist:2001mj},  origin of electroweak symmetry breaking \cite{ArkaniHamed:2000hv,Hashimoto:2000uk,Csaki:2002ur,Scrucca:2003ut}, breaking of grand unified gauge groups \cite{Hebecker:2001jb,Hall:2001xr,Asaka:2002my,Asaka:2003iy} and  the number of fermion
generations \cite{Dobrescu:2001ae,Fabbrichesi:2001fx,Borghini:2001sa,Fabbrichesi:2002am,Frere:2001ug,Watari:2002tf}. Many extensions of the SM appear by employing ED as well; indeed even the SM itself can be embedded in ED, whose  fields propagate
in the compact ED. In 4-D, the zero mode of each Kaluza-Klein (KK) tower of states is  identified with the correspondent SM particle. These so-called  Universal Extra 
Dimension (UED) models were build either with one \cite{Appelquist:2000nn} or two ED \cite{Dobrescu:2004zi,Burdman:2005sr,Ponton:2005kx,Burdman:2006gy}, for example, and current results from LHC \cite{Aad:2015mia,TheATLAScollaboration:2013uha} impose bounds on the UED compactification radius $L$ for one  ($L^{-1}>1.4-1.5$ TeV) 
\cite{Deutschmann:2017bth,Beuria:2017jez,Tanabashi:2018oca} (for $\Lambda L \sim 5-35$, where $\Lambda$ is
the cutoff scale) or two ED ($L^{-1}>900$ GeV) \cite{Burdman:2016njl}.  
 In the context of ED, the DP model was embedded in a flat, single ED, along with  DM candidates \cite{Rizzo:2018ntg,Rizzo:2018joy}.

Much of the parameter space for the kinetic-mixing term  has been excluded by several experiments and observations \cite{Aartsen:2012kia,Aartsen:2016exj,Battaglieri:2017aum,Riordan:1987aw,Bjorken:1988as,Bross:1989mp,Bjorken:2009mm, Pospelov:2008zw,Davoudiasl:2012ig,Endo:2012hp,Babusci:2012cr,Archilli:2011zc,Adlarson:2013eza,Abrahamyan:2011gv,Merkel:2011ze,Reece:2009un,Kazanas:2014mca,Chang:2016ntp}. Its expected small value  may be explained if one considers a single, flat ED and a thick brane \cite{Landim:2019epv}, where the presence of a brane-localized-kinetic term (BLKT) spread inside the fat brane increase the suppression mechanism. BLKT appears as loop corrections associated with localized matter fields, giving rise to a massless spin-2 field \cite{Dvali:2000rx} or massless spin-1 field \cite{Dvali:2000hr}. The  same mechanism also works  for two ED  \cite{Dvali:2000xg,Dvali:2001ae,Dvali:2002pe}, where the induced kinetic term is effectively 4-D, meaning that any expected extra scalar field, arising from the compactification of the ED has no contribution in 4-D. Thus for the graviton, for instance,  the induced term on the brane describes a 4-D tensor gravity, rather than a 4-D tensor-scalar gravity.    The role of BLKT has been investigated in several different scenarios \cite{Carena:2002me,Carena:2002dz,delAguila:2003gv,delAguila:2003bh,Davoudiasl:2002ua,Davoudiasl:2003zt, Freitas:2017afm,Flacke:2017xsv,Flacke:2013pla,Gao:2014wga,Flacke:2014jwa,Flacke:2013nta}, while the localization of matter or gauge fields in branes has been studied in other contexts, for thin \cite{ArkaniHamed:1998rs,Dvali:1998pa, Alencar:2014moa,Alencar:2017dqb,Alencar:2015awa,Alencar:2015rtc,Alencar:2015oka,Alencar:2018cbk,Freitas:2018iil} and thick branes  \cite{DeRujula:2000he,Georgi:2000wb}.

 In a recent paper \cite{Landim:2019ufg}, a model similar to the one presented in \cite{Landim:2019epv} was explored in 6-D, however it was shown that it is not possible to have a BLKT inside the fat brane, since the wave-functions do not satisfy the boundary conditions (BC) all along the boundary. Although it is expected to have BLKT in both thin and thick branes, one may investigate the case where the BLKT inside the fat brane is very small and can be neglected. This is the aim of this paper, where we show that it is possible to have a vector field in the bulk, which mediates the interaction between the SM particles localized in  the fat brane  and a DM candidate confined in a thin brane, without employing a kinetic-mixing term. The coupling with the SM, although not as suppressed as in \cite{Landim:2019epv}, has a similar behavior. In this
framework we can obtain the observed  DM relic abundance for a  range of
parameter choices, as well as avoid DM direct detection constraints.

This paper is organized as follows. In Sect. \ref{General Framework Setup} we present vector mediate, a  6-D gauge 
field with BLKT on the chiral square. In Sect. \ref{gauge-DM}  we analyze
the resulting couplings with the SM and DM through the vector mediator. We examine the constraints on the SM
interactions with the DM particle from both direct and indirect observations in Sect, \ref{sec:constraints}, while  Sect. \ref{sec:discussion} is reserved for conclusions.

\section{ Vector mediator in the bulk}\label{General Framework Setup}

We will consider two flat and transverse ED ($x^4$ and $x^5$) compactified on the chiral square. The chiral square is chosen in the UED model with two ED because it is the simplest compactification that leads to chiral quarks and leptons in 4-D \cite{Dobrescu:2004zi}. The square has size $\pi R$, where $R$ is the compactification radius of the ED,  and the adjacent sides are identified $(0,y)\sim (y,0)$ and $(\pi R,y)\sim (y,\pi R)$, with $y\in [0,\pi R]$. This means that the Lagrangians at those points have the same values for any field configuration: $\mathcal{L}(x^\mu, 0,y)=$ $\mathcal{L}(x^\mu,y,0)$ and $\mathcal{L}(x^\mu, \pi R,y)=\mathcal{L}(x^\mu,y,\pi R)$. A thin brane is localized at the origin $(0,0)$, where the DM candidate is confined, and the  SM is contained within a fat
brane, lying between $(\pi r, \pi r)$ and $(\pi R, \pi R)$, with a width $\pi(R-r)\equiv \pi L$, such that we assume  $L\ll R$. The radius $r$ represents the amount of the ED that is not part of the thick brane. 

There is one abelian gauge field $V^A, ~A=0-3, 4, 5$ in the bulk, interacting  both with  DM and SM. Since we are not assuming kinetic mixing, the vector field couples with DM and SM 
through the covariant derivative, which contains  a term proportional to $\sim g_{6D}(B-L+Q_D)$, where $g_{6D}$ is  the 6-D dark gauge coupling.  SM particles have $B-L\neq 0$ and $Q_D=0$, while  
DM has $B-L=0$ by assumption and, without loss of generality,  $Q_D=1$.  We will use one of the four anomaly-free 
symmetries  which do not need any additional SM
fermion fields (beyond right-handed neutrinos, \textit{i.e.}, the difference between baryon and lepton numbers ($U_{B-L}$) and the three differences between the lepton numbers 
($U_{L_\mu-L_e}$, $U_{L_e-L_\tau}$ and $U_{L_\mu-L_\tau}$ \cite{Foot:1990mn,He:1990pn,He:1991qd,Bauer:2018onh}), under which only
baryons and/or leptons are charged. This is done in order to avoid dangerous couplings with the Higgs or gauge bosons, which in turn would spoil  some of the well constrained electroweak predictions  \cite{Tanabashi:2018oca}, such as the Z boson mass.

The action is similar to the one of UED model with two ED \cite{Dobrescu:2004zi,Burdman:2005sr}, given by
\begin{align}\label{eq:actionV}
    S=&\int d^4x\int_0^{\pi R} dx^4\int_0^{\pi R}dx^5 \Big( -\frac{1}{4}V_{AB}V^{AB}+\mathcal{L}_{GF}\nonumber\\&+\mathcal{L}_{BLKT} \Big)\,,
\end{align}
where A is the 6-D index and the gauge fixing term has the following form to cancel the mixing between $V_4$ and $V_5$ with $V_\mu$ \cite{Burdman:2005sr}
\begin{equation}\label{eq:actionGF}
\mathcal{L}_{GF}=-\frac{1}{2}\Big[\partial_\mu V^\mu-(\partial_4V_4+\partial_5V_5)\Big]^2\,,
\end{equation}
where  we will work in the Feynman gauge.
We will consider BLKT  at the point $(0,0)$, where is localized the thin brane. Any BLKT on the fat brane should be very small \cite{Landim:2019ufg} and will be neglected. Notice that KK parity is not preserved, although usually one invokes this $Z^2_{KK}$ symmetry in UED models in order for the lowest KK
state to stable and to be the DM candidate, which  is not needed in our case because the DM candidate is confined on the thin brane.   

The BLKT at $(0,0)$ contributes with a term
\cite{Dvali:2000rx,Dvali:2001ae}
\begin{equation}\label{eq:lagrangianBLKTthin}
\mathcal{L}_{BLKT}=\left[-\frac{1}{4}V_{\mu\nu}V^{\mu\nu}-\frac{1}{2}(\partial_\mu V^\mu)^2\right]\cdot \delta_A R^2\,\delta(x^4,x^5)\,,
\end{equation}
where $\delta_A$ is positive constant.

Expanding the components of the 6-D gauge field in KK towers of states
  \begin{equation}\label{eq:KKexpAmu}
      V_\mu(x^\nu, x^4,x^5)=\sum_j\sum_k v_0^{(j,k)}(x^4,x^5) V_\mu^{(j,k)}(x^\nu)\,,
  \end{equation}

  \begin{equation}\label{eq:KKexpA4}
      V_4(x^\nu, x^4,x^5)=\sum_j\sum_k v_4^{(j,k)}(x^4,x^5) V_4^{(j,k)}(x^\nu)\,,
  \end{equation}
  
  \begin{equation}\label{eq:KKexpA5}
      V_5(x^\nu, x^4,x^5)=\sum_j\sum_k v_5^{(j,k)}(x^4,x^5) V_5^{(j,k)}(x^\nu)\,,
  \end{equation}
 leads to the solutions of the equations of motion for $v_4^{(j,k)}(x^4,x^5)$ and $v_5^{(j,k)}(x^4,x^5)$  \cite{Burdman:2005sr}\footnote{ In \cite{Burdman:2005sr} the authors made the linear combinations $V_{\pm}=V_4\pm i V_5$.}
    \begin{equation}\label{eq:v4}
     v_4^{(j,k)}(x^4,x^5)=\frac{\sqrt{2}}{\pi R}\sin \Big(\frac{j x^4+k x^5}{R}\Big)\,,
  \end{equation}
      \begin{equation}\label{eq:v5}
     v_5^{(j,k)}(x^4,x^5)=-\frac{\sqrt{2}}{\pi R}\sin \Big(\frac{k x^4-j x^5}{R}\Big)\,,
  \end{equation}
  where $j$ and $k$ are integers. 
The physical masses  of these scalar fields are $ (M_{4,5}^{(j,k)})^2=(j^2+k^2)/R^2$, and notice that $V_4=V_5=0$ for $j=k=0$, from Eqs. (\ref{eq:v4}) and (\ref{eq:v5}). Notice that because the scalar fields vanish at the thin brane, they do not interact with the  DM (lying also in the thin brane).

The equation of motion for the wave-function $  v_0^{(j,k)}(x^4,x^5)$ is
\begin{equation}\label{eq:vx4x5delta}
     \Big[\partial_4^2+\partial_5^2+M_{j,k}^2+ M_{j,k}^2\delta_AR^2\delta(x_4,x_5)\Big]v_0^{(j,k)}(x^4,x^5)=0\,,
  \end{equation}
  where 
  \begin{equation}\label{mass}
   M_{j,k}^2=m_j^2+m_k^2\,,
\end{equation}
whose solution yields \cite{Landim:2019ufg}
  \begin{align}\label{eq:v0deltafunction}
      v_0^{(j,k)}(x^4,x^5)=&N_{j,k}\Big[\cos (m_j x^4)\cos (m_k x^5)\nonumber\\
    &+\cos (m_k x^4)\cos (m_j x^5)\nonumber\\&-\frac{\delta_A}{2}x_j x_k\Big (\sin (m_j x^4)\sin (m_k x^5)\nonumber\\
    &+\sin (m_k x^4)\sin (m_j x^5)\Big)\Big]\,,
  \end{align}
  where $m_j=x_j/R$ and $m_k=x_k/R$.   The normalization constant $N_{j,k}$ is defined through
   \begin{eqnarray}\label{eq:normalizationcondit}
     \int_0^{\pi R} dx^4\int_0^{\pi R} dx^5\, v_i^{(j,k)}(x^4,x^5)v_i^{(j',k')}(x^4,x^5)&=\delta_{j,j'}\delta_{k,k'}\,,
\end{eqnarray}
which results for $j\neq k$ in
  \begin{align}\label{Nmn}
      N_{j,k}^{-2}=&\frac{\pi ^2R^2}{2}\biggr \{1 +\frac{\delta_A}{4\pi ^2}  \cos ^2(\pi  x_j)\Big[1+\cos ^2(\pi  x_k)\Big]+\frac{1}{4}  \delta_A^2 x_j^2 x_k^2\nonumber\\
   &-\frac{ \delta_A}{2\pi}\Big[ x_k \cos ^2(\pi  x_j) \cot(\pi  x_k)+ x_j \cot (\pi  x_j) \cos ^2(\pi  x_k)\Big]\nonumber\\
    &-\frac{x_j x_k \sin(2 \pi  x_j) \sin (2 \pi  x_k)}{ \pi^2(x_j^2-x_k^2){}^2}+\frac{4 x_k^2 \cos ^2(\pi  x_j) \csc ^2(\pi  x_k)}{\pi^2(x_j^2-x_k^2){}^2}\nonumber\\&+\frac{4 x_j^2 \csc ^2(\pi  x_j) \cos ^2(\pi  x_k)}{\pi^2(x_j^2-x_k^2){}^2} +\frac{  \sin (2 \pi  x_k)}{2 \pi x_k}\biggr\}\,,
  \end{align}
  and for $j=k$ in
  \begin{align}\label{Nm}
     N_{j=k}^{-2}=& \pi ^2 R^2\biggr\{1+ \frac{\delta_A^2x_j^4}{4\pi^2}   +\frac{\delta_A^2x_j^2 }{16\pi^2}   \sin ^2\left(2 \pi  x_j\right)-\frac{\delta_A^2x_j^3}{4\pi}  \sin \left(2 \pi  x_j\right)\nonumber\\&-\frac{\delta_A}{\pi^2}\sin ^4\left(\pi  x_j\right)+\frac{ \sin ^2\left(2 \pi  x_j\right)}{4 \pi^2x_j^2}+\frac{ \sin \left(2 \pi  x_j\right)}{\pi x_j}\biggr\}\,.
  \end{align}
At a first glance one might think that there are imaginary values for the normalization constants above, however the allowed values of $x_j$ and $x_k$ do not lead to complex numbers in Eqs. (\ref{Nmn}) and (\ref{Nm}).

 The 4-D gauge field is canonically normalized through the relations  \begin{align}\label{eq:normalizationconditdelta}
     &\int_0^{\pi R} dx^4\int_0^{\pi R} dx^5\,\Big[1+\delta_AR^2\delta(x^4,x^5)\Big] v_0^{(j,k)}v_0^{(j',k')}\nonumber\\
     &=Z_{(j,k)}\delta_{j,j'}\delta_{k,k'}\,,\nonumber\\
     & \int_0^{\pi R} dx^4\int_0^{\pi R} dx^5\,\Big[\partial_4v_0^{(j,k)}\partial_4v_0^{(j',k')}+\partial_5v_0^{(j,k)}\partial_5v_0^{(j',k')}\Big]\nonumber\\
    &=Z_{(j,k)}M_{j,k}^2\delta_{j,j'}\delta_{k,k'} \,,
\end{align}
where $Z_{(j,k)}$ is a normalization factor $ Z_{(j,k)}=1+\delta_A R^2v_0^{(j,k)}(0,0)\,,$
 
 The transcendental equation that determines the roots $x_j$ and $x_k$ is found requiring the Dirichlet BC $ v_0^{(j,k)}(\pi R,\pi R)=0$, whose solutions depend only upon the parameter $\delta_A$
\begin{equation}\label{eq:root}
    \cot(\pi  x_j)\cot(\pi  x_k)=\frac{\delta_A}{2}x_j x_k\,.
\end{equation}

\begin{figure}
    \centering
    \includegraphics[scale=0.4]{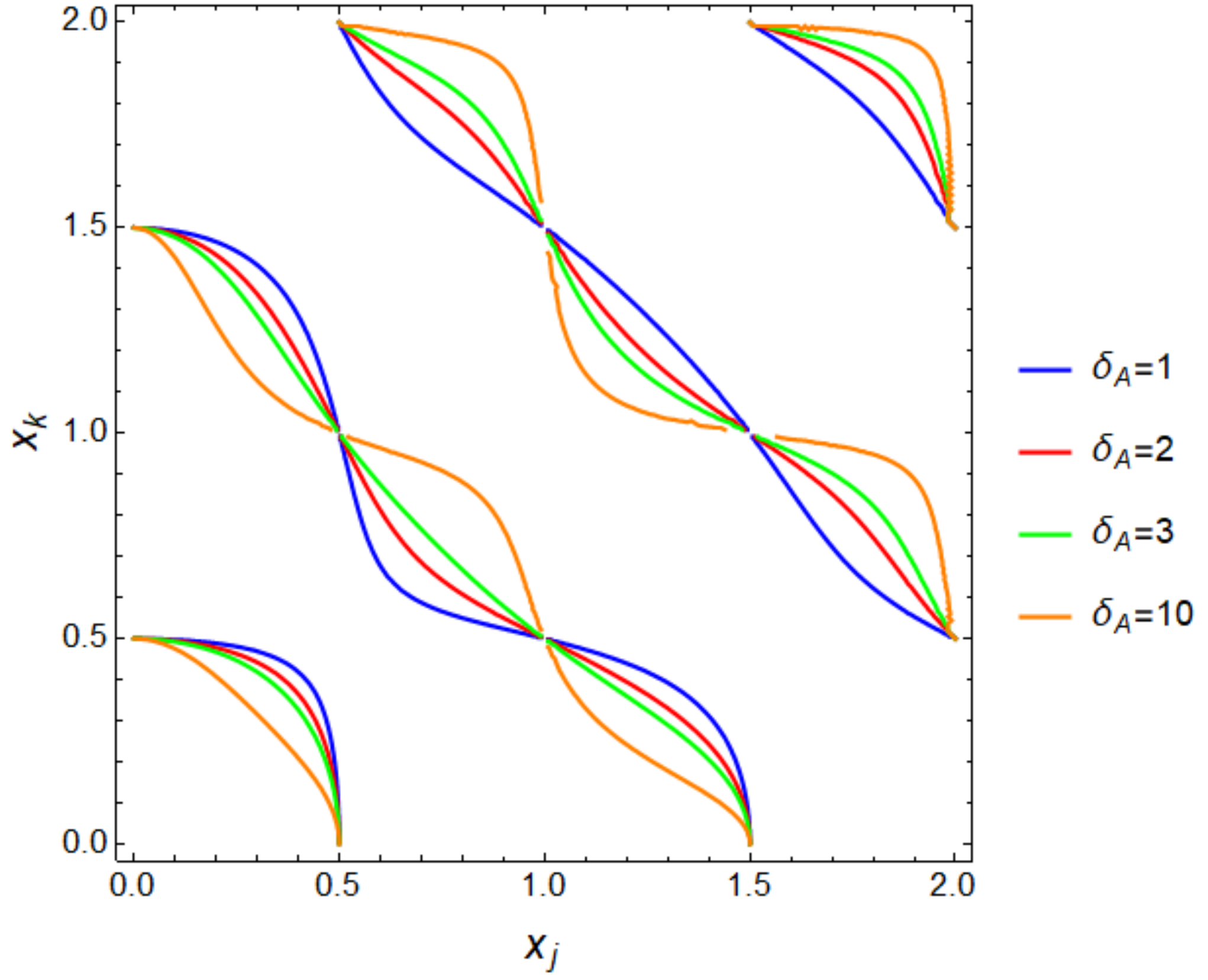}
    \caption{Solutions of the transcendental equation  (\ref{eq:root}) for different values of $\delta_A$.}
    \label{fig:rootEqA}
\end{figure}

 The solutions of Eq.  (\ref{eq:root}), given in \cite{Landim:2019ufg}, are reproduced in Fig. \ref{fig:rootEqA}, for different values of $\delta_A$. There are $(2n+1)$ quantized masses for each curve  $n$, where  $n$  is each one of the dashed lines. Each mode is described by the segments in the dashed lines, thus there is one mode for $n=0$, a massive zero-mode $M_{0,0}$, while the second dashed line ($n=1$) has three  quantized masses $M_{0,1}$, $M_{1,0}$ and $M_{1,1}$, being the first two degenerate, etc. Notice that the masses $M_{j,k}$ and $M_{k,j}$ are degenerate.  The whole continuous set of values $(x_j, x_k)$  in each segment represent only one  mass state, being narrow the range of each state \cite{Landim:2019ufg}.

\section{Interactions }\label{gauge-DM}
The couplings between the tower of (KK) mediators and DM is  $g_{D,(j,k)}= g_{D} N_{j,k}/N_{0,0}  $, where $g_D\equiv g_{6D}N_{0,0}$, $g_{6D}$ is the 6-D  dark gauge coupling and $N_{0,0}$ is the normalization of the lowest KK state ($j=k=0$). 
On the other hand, the interaction between the vector field and a generic  zeroth-mode SM field $\phi$, localized inside the fat brane, is given by the integral over the brane width
\begin{equation}
    \int_{\pi r}^{\pi R}dx^4 \int_{\pi r}^{\pi R}dx^5\, V^{\mu}J_{\mu}\,,
\end{equation}
where $J_\mu$ is the SM current. We are interested in the interaction with   conventional SM particles, thus the zeroth-mode of  SM field  in the 6D UED model is 
$\phi/(\pi^2 L^2)$, where $(\pi L)^{-1}$ is the normalization constant. The 4-D gauge couplings between the SM fields and the KK mediators are defined as
\begin{align}\label{eq:gauge-coupling-full}
    g_{D,(j,k)}^{ED}&\equiv g_{6D} \int_{\pi r}^{\pi R}dx^4 \int_{\pi r}^{\pi R}dx^5\, \frac{v_{0}^{(j,k)}(x^4,x^5)}{\pi^2 L^2}\nonumber\\
    \end{align}
    which in turn yields 
    \begin{align}\label{gaugecouplingjk}
     g_{D,(j,k)}^{ED}&=\frac{g_{D} N_{j,k}R^2 }{N_{0,0}\pi^2 L^2}  \biggr\{\frac{2}{x_j x_k} \left[\sin \left(\pi  x_j\right)-\sin \left(\frac{\pi  r x_j}{R}\right)\right]\nonumber\\&\times \left[\sin \left(\pi  x_k\right)-\sin \left(\frac{\pi  r x_k}{R}\right)\right]\nonumber\\&-\delta_A \left[\cos \left(\pi  x_j\right)-\cos \left(\frac{\pi  r x_j}{R}\right)\right]\nonumber\\&\times \left[\cos \left(\pi  x_k\right)-\cos \left(\frac{\pi  r x_k}{R}\right)\right]\biggr\}\,,
\end{align}
for $j\neq k$, while for $j=k$ the result is
\begin{align}\label{gaugecouplingjj}
     g_{D,(j,j)}^{ED}&=\frac{g_{D} N_{j,j}2R^2 }{N_{0,0}\pi^2 L^2 x_j^2}  \biggr\{ \sin ^2\left(\frac{\pi  x_j L}{2 R}\right)\nonumber\\&\times \left(\left(\delta_A x_j^2+2\right) \cos \left(\frac{\pi  x_j (r+R)}{R}\right)-\delta_A x_j^2+2\right)\biggr\}.
\end{align}
From Eqs. (\ref{gaugecouplingjk}) and specially (\ref{gaugecouplingjj}) we see that in the limit of small roots $x_j$, since $L\ll R$, $ g_{D,(j,j)}^{ED}\sim \frac{g_{D} N_{j,j} }{2N_{0,0}}$,  while in 5-D the coupling is reduced by a factor proportional to $L/R$ \cite{Landim:2019epv}. This behavior does not occur here because there is no significant BLKT in the fat brane. 

For illustrative purposes,  we will consider four specific benchmark models (BM), whose assumed set of values for the compactification radius $R$ and the width of the fat brane $L$ are shown in Table \ref{table:param}.
 
\begin{table}[h] 
       \centering 
           \begin{tabular}{c c c c c}
           \hline
    BM  &   I &  II &  III &  IV\\
   \hline
    $R^{-1} $  & $1 $ GeV &  $1$ GeV  & $100$  MeV & $100$ MeV  \\
    $L^{-1}$  &$1$ TeV  & $10$ TeV  & $1$ TeV & $10$ TeV  \\
         \hline \end{tabular}
 \caption{Illustrative sets of compactification radius $R$ and fat brane width $L$.}   \label{table:param}
\end{table}

\begin{figure*}
    \centering
    \includegraphics[scale=0.45]{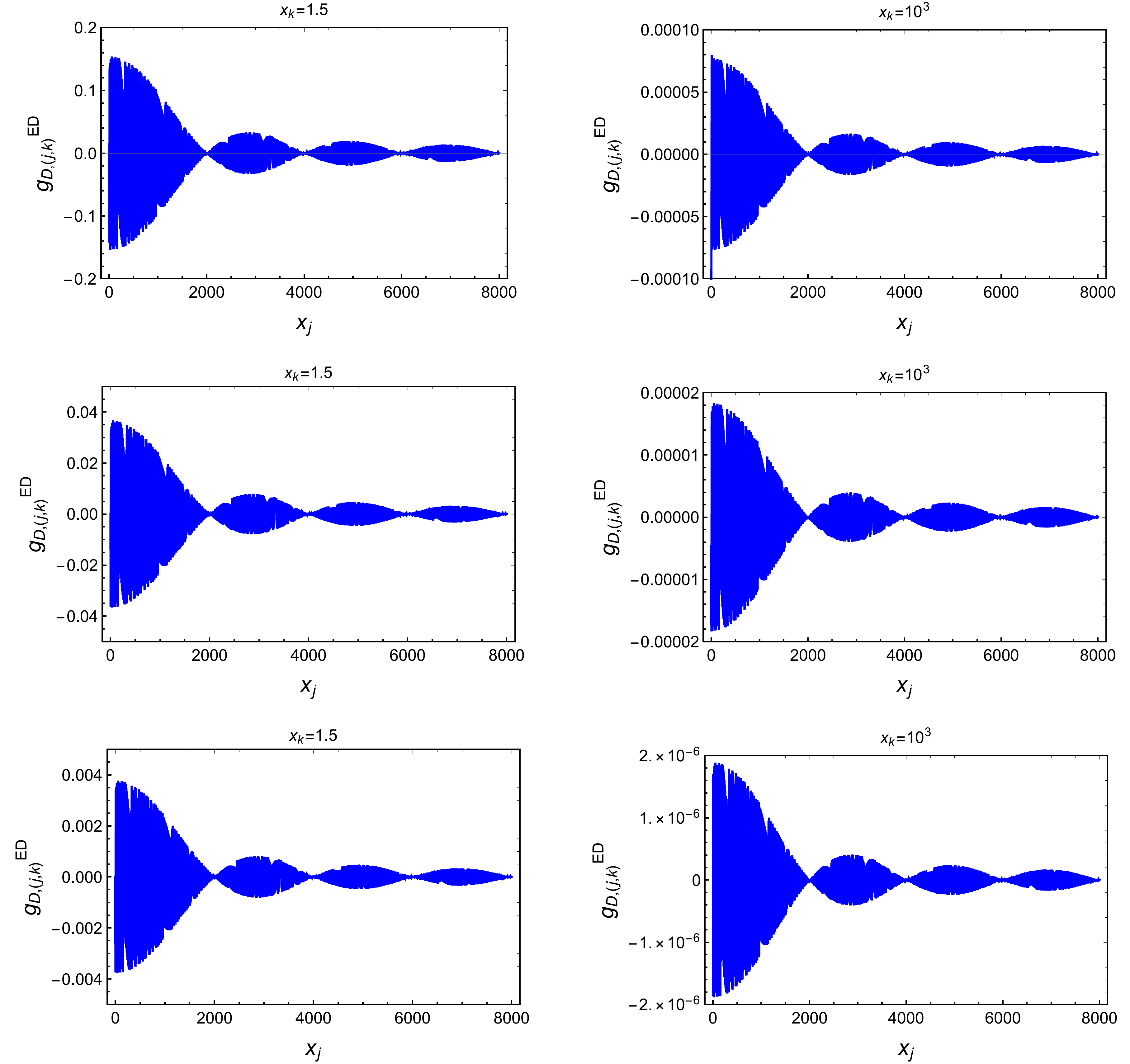}
    \caption{Gauge KK couplings as a function of $x_j$  with $x_k $ fixed at around $x_k\sim 1.5$ (left) and $x_k\sim 10^3$ (right), for $g_{D}=1$, $ \delta_A=1 $ (first row), $\delta_A=10 $ (second row) and $\delta_A=100 $ (third row), for the BM I.}
    \label{fig:plotBM1}
\end{figure*}

\begin{figure*}
    \centering
    \includegraphics[scale=0.45]{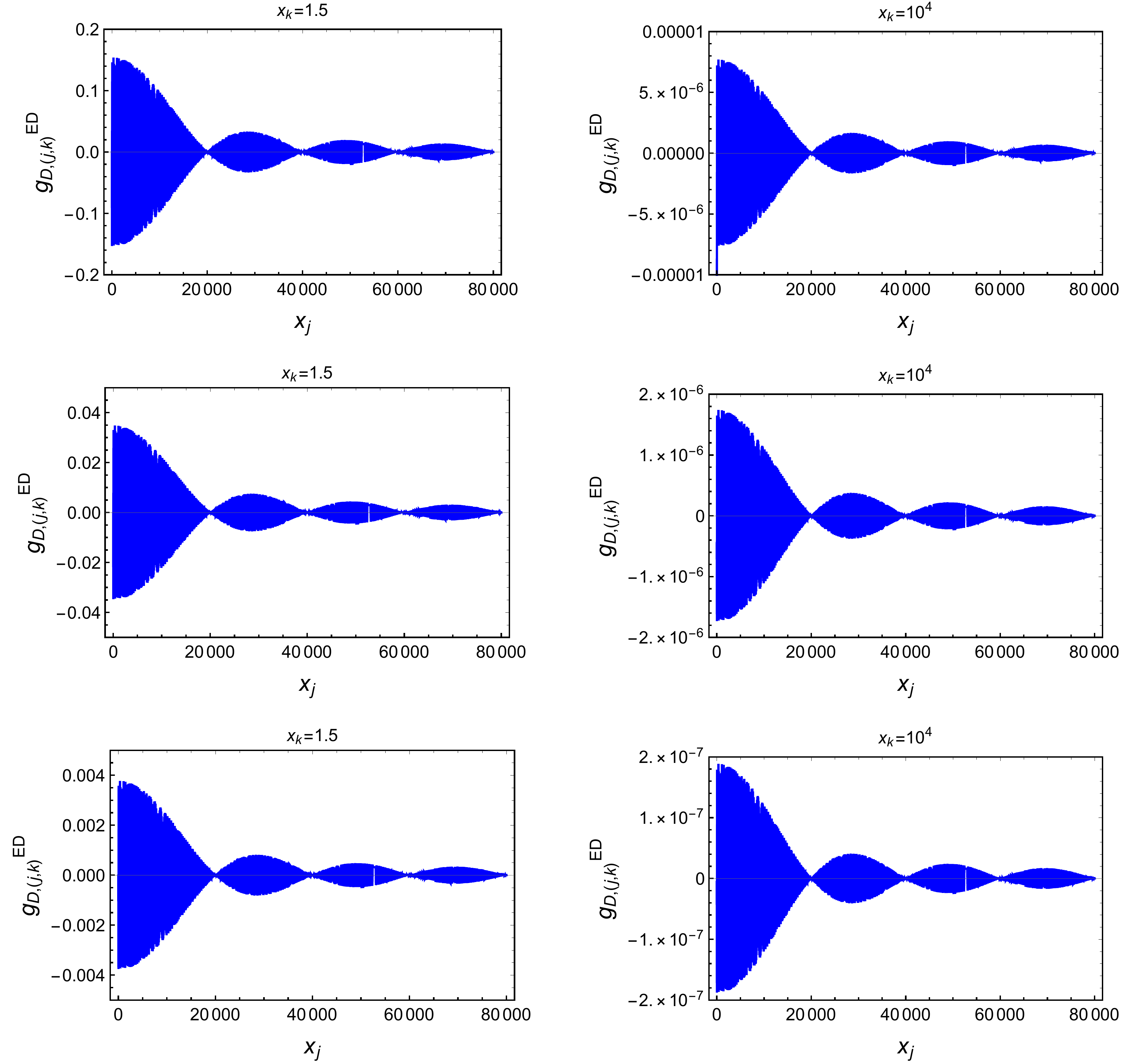}
    \caption{Gauge KK couplings as a function of $x_j$  with $x_k $ fixed at around $x_k\sim 1.5$ (left) and $x_k\sim 10^4$ (right), for $g_{D}=1$, $ \delta_A=1 $ (first row), $\delta_A=10 $ (second row) and $\delta_A=100 $ (third row), for the BM II.}
    \label{fig:plotBM2}
\end{figure*}

\begin{figure*}
    \centering
    \includegraphics[scale=0.45]{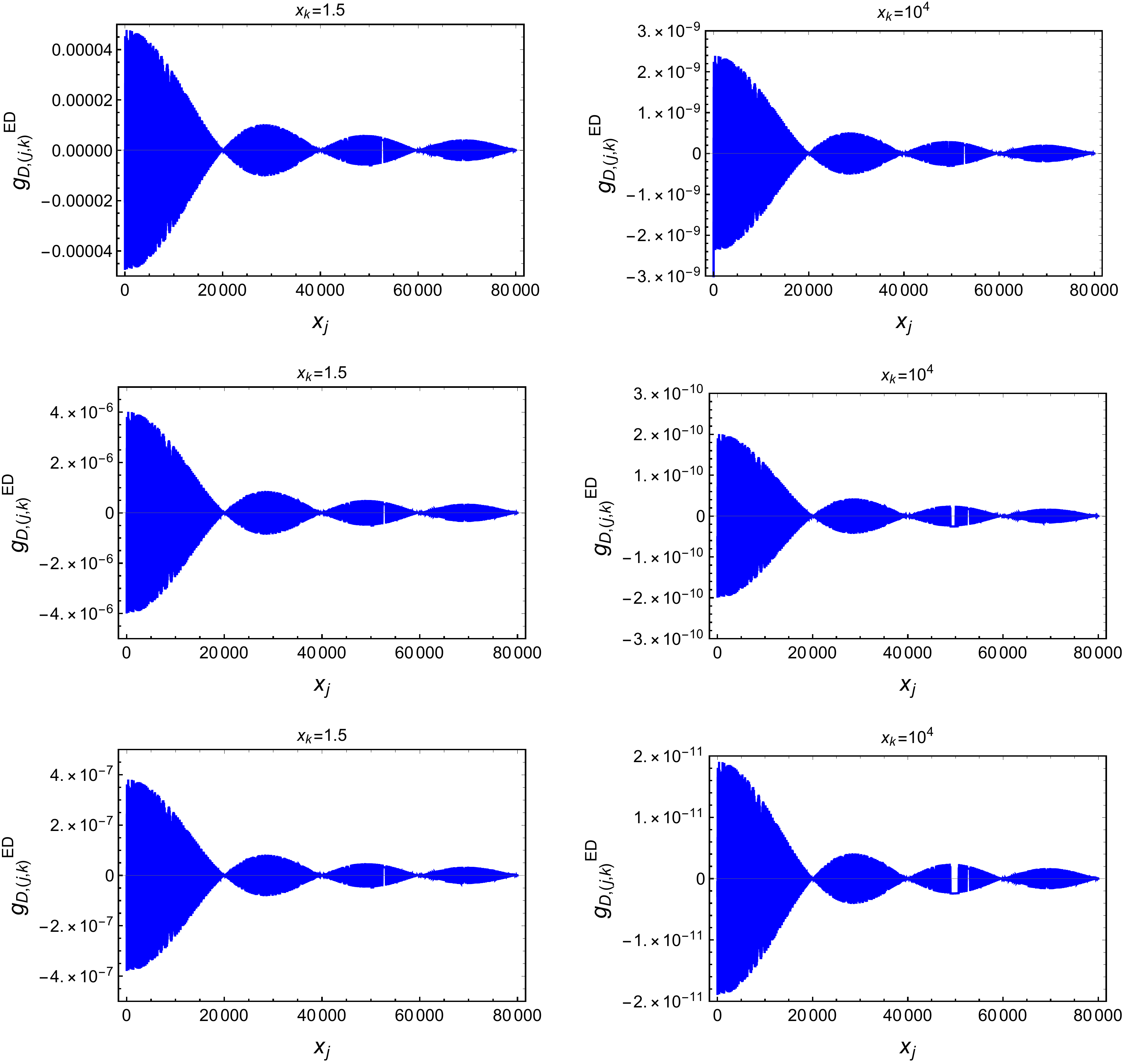}
    \caption{Gauge KK couplings as a function of $x_j$  with $x_k $ fixed at around $x_k\sim 1.5$ (left) and $x_k\sim 10^4$ (right), for $g_{D}=1$, $ \delta_A=1 $ (first row), $\delta_A=10 $ (second row) and $\delta_A=100 $ (third row), for the BM III.}
    \label{fig:plotBM3}
\end{figure*}

\begin{figure*}
    \centering
    \includegraphics[scale=0.4]{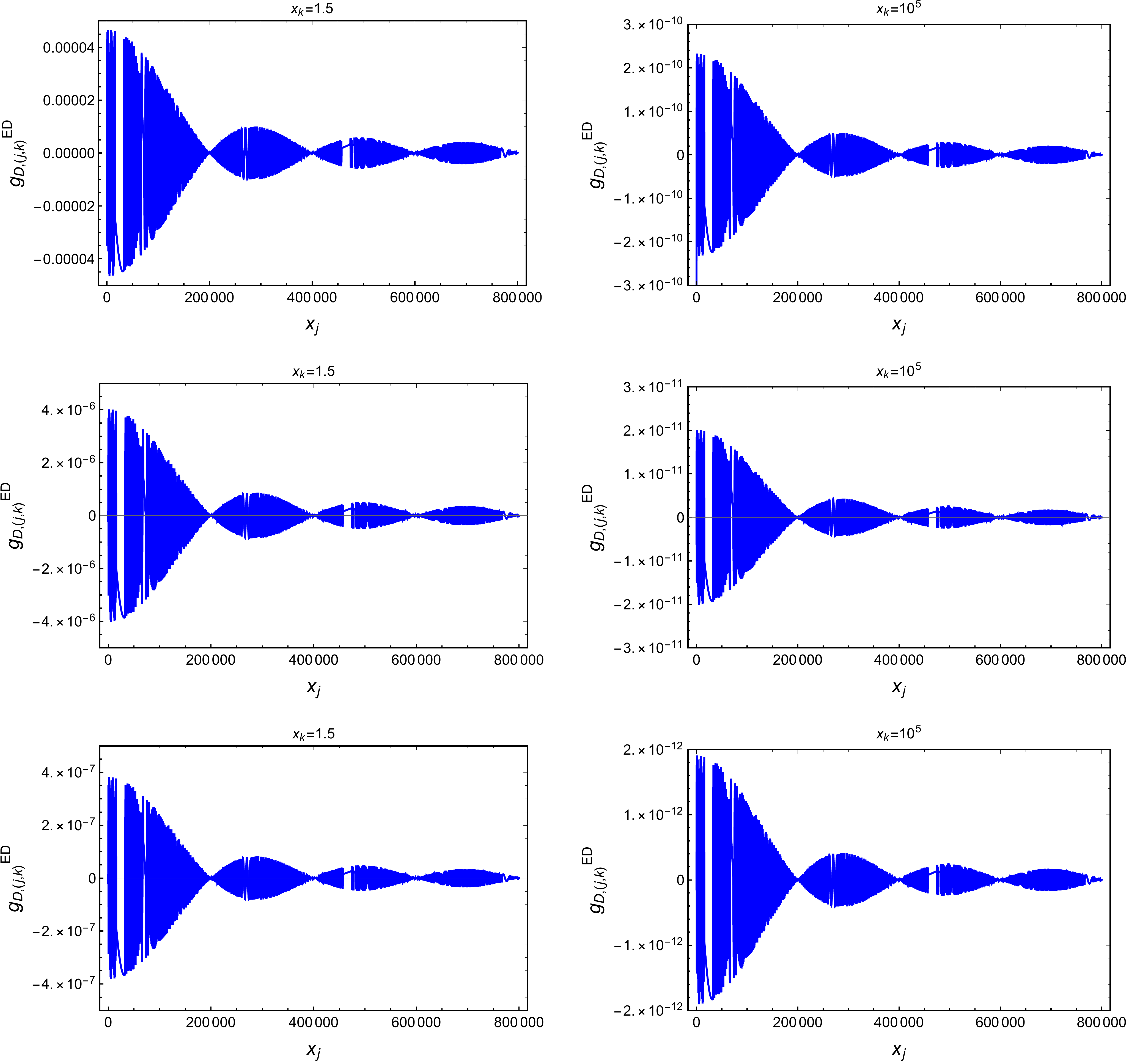}
    \caption{Gauge KK couplings as a function of $x_j$  with $x_k $ fixed at around $x_k\sim 1.5$ (left) and $x_k\sim 10^5$ (right), for $g_{D}=1$, $ \delta_A=1 $ (first row), $\delta_A=10 $ (second row) and $\delta_A=100 $ (third row), for the BM IV.}
    \label{fig:plotBM4}
\end{figure*}

In Figs. \ref{fig:plotBM1}--\ref{fig:plotBM4} we plot the oscillatory behavior of the gauge coupling (\ref{gaugecouplingjk}),  for different values of $\delta_A$ and for the four BM in Table \ref{table:param}. The coupling decreases as $x_j$ and $x_k$ increases, and although it maintains the same pattern for different values of $\delta_A$, the coupling is orders of magnitude smaller  as $\delta_A$ is increased. The difference between the four BM is similar to the one presented in \cite{Landim:2019epv}: all BM show the same oscillatory pattern,  but BM II  (IV) reproduce the exact plot in  II (III) after ten times more roots $x_j$, while decreasing the inverse of the compactification radius $R$ makes the coupling smaller (compare BM II in Fig. \ref{fig:plotBM2} with III in Fig. \ref{fig:plotBM3}).

\section{Constraints from observations and experiments}\label{sec:constraints}

We now consider the interactions between the DM candidate confined at the origin of the chiral square with the SM, mediated by the KK tower of states. We assume a complex scalar field as a DM candidate for simplicity, which is naturally stable via the $Z_2$ symmetry. The couplings between the KK mediators and the DM is  $g_{D,(j,k)}\equiv g_D N_{j,k}/N_{0,0}$, while inside the fat brane the coupling is $g_{D,{(j,k)}}^{ED}$ as described in Eqs. (\ref{gaugecouplingjk}) and (\ref{gaugecouplingjj}). The mass and couplings of the DM particle are constrained by both direct and indirect experiments. In order for the DM not to annihilate into a pair of mediator particles (avoiding the  s-wave annihilation  excluded by Planck results \cite{Aghanim:2018eyx}), DM should be lighter than the lightest mediator KK state. Thus the 
DM mass must be smaller than the lowest mediator mass  $x_{0,0}/R$, whose root $x_{0,0}$ lies  in the range $\sim 0.4-0.5$. 

Assuming that the  DM relic abundance is due to thermal freeze-out,  the resulting final states from DM pair annihilation can be $e^+e^-$ and $\mu^+\mu^-$,   as well as three generations of nearly massless neutrinos, given the DM mass range of interest and only the final states that are charged under $B-L$. Note that the only  accessible channels for BM II and IV are $e^+e^-$ and neutrinos, because DM particle is lighter than muons. Considering the usual expansion of the thermally- averaged cross section (away from the resonance)  in powers of the relative velocity of  DM particles, 
$v^2$, given by $\sigma v \approx a+bv^2$, we have the following coefficients for  a vector mediator and a complex scalar DM:  $a=0$ and
  \cite{Berlin:2014tja}
\begin{equation}\label{eq:b}
   b_f=\frac{m_{DM}^2}{ 6 \pi }\sqrt{1-\frac{m_f^2}{m_{DM}^2}}\left(1-\frac{m_f^2}{2m_{DM}^2}\right)\left(\sum_n\frac{g_{D,(j,k)} g_{D,(j,k)}^{ED}}{M_{j,k}^2-4m_{DM}^2}\right)^2\,,
\end{equation}
where $m_f$ is the mass of the final states.  The (dominant) p-wave DM annihilation is therefore not constrained by current observations \cite{Aghanim:2018eyx,Leane:2018kjk}.

The observed value of the DM relic density  is obtained then through   \cite{Berlin:2014tja} 
\begin{equation}
    \Omega h^2 \simeq \frac{x_f\, 1.07\times 10^9  \,\text{GeV}^{-1}}{g_*^{1/2}M_{Pl}(a+3b/x_f)}\,,
\end{equation}
where $M_{Pl}$ is the Planck mass and $x_f\equiv m_{DM}/T_f$ is the usual ratio between the DM mass and the temperature at the freeze-out, which can be taken to be $x_f=20$. 
The effective number of degrees of freedom for the range of DM masses of interest here ($40-700$ MeV) is $g_*\simeq 10.75$, since the temperature at the freeze-out 
is $\sim 2-40$ MeV. The value of the coupling $g_D$ which gives the observed relic density ( $\Omega h^2=0.12$ \cite{Aghanim:2018eyx}) is calculated for the four BM with different values of $\delta_A=0.1,1,10$. The lightest KK mediator  mass is  presented in Table \ref{tab:relicDens}, for these parameter choices. As we can see  the mass of the lightest KK state increases as $\delta_A$ decreases.
 
 \begin{figure}
     \centering
     \includegraphics[scale=0.5]{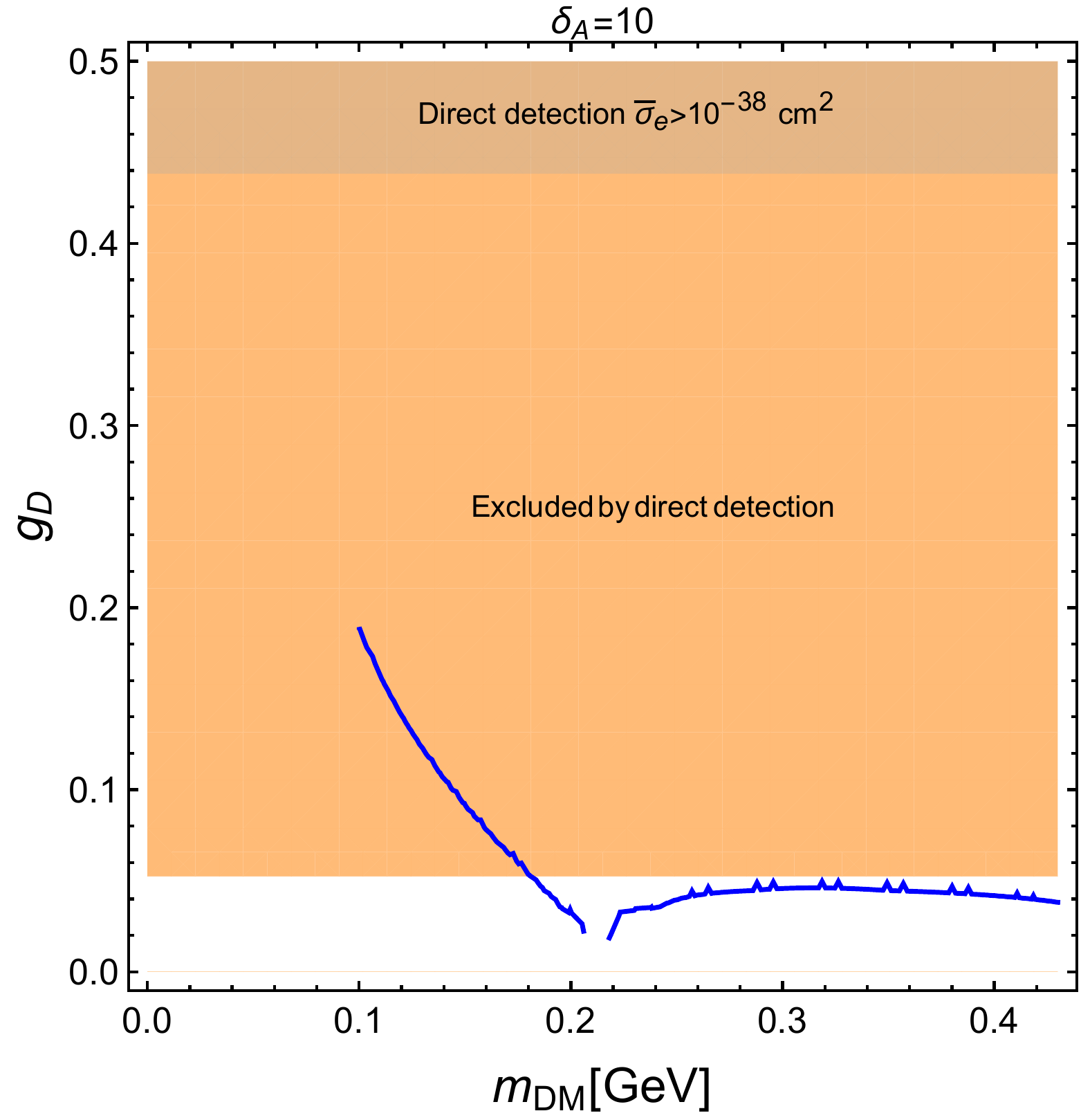}
     \caption{Allowed parameter space (BM II) that can satisfy the observed DM relic abundance (blue line). Notice that $m_{DM}<M_{0,0}$. The region in orange and brown are excluded by current direct detection experiments. As shown in Eq. (\ref{eq:b}) DM masses were assumed, for simplicity, to be away from the resonant region. }
     \label{fig:plotlimitBM2delta10}
 \end{figure}

\begin{table}[h] 
       \centering 
           \begin{tabular}{c c c c c}
           \hline
    BM  &   I &  II &  III &  IV\\
    
   \hline
   $\delta_A=10^{-1}$ & & & &\\
          \hline
           $M_{0,0}$ [MeV] &710 &  710 & 71 & 71  \\

         \hline
   $\delta_A=1$ & & & &\\
   \hline
    $M_{0,0}$ [MeV] &570 &  570 & 57 & 57  \\
  
         \hline
          
         $\delta_A=10$ & & & &\\
          \hline
            $M_{0,0}$ [MeV] &420 &  420 & 42 & 42  \\
  
            \hline \end{tabular}
 \caption{Lightest vector mediator mass $M_{0,0}$ for $ \delta_A=10^{-1},1 \text{ or } 10$.  }   \label{tab:relicDens}
\end{table}

Since  DM masses here are relatively light and the corresponding recoil energies in direct detection is  small, DM scattering off electrons provides greater sensitivity \cite{Battaglieri:2017aum}. In our case the scattering cross section then becomes \cite{Essig:2011nj,Berlin:2014tja,Emken:2017erx}
\begin{equation}\label{sigmae}
    \sigma_e=\frac{\mu^2}{4   \pi  }\left(\sum_{j,k}\frac{g_{D,(j,k)} g_{D,(j,k)}^{ED}}{M_{j,k}^2}\right)^2\,,
\end{equation}
where a form factor of unity has been assumed and the reduced mass $\mu=m_em_{DM}/(m_e+m_{DM})\sim m_e$, since $m_{DM}^2\gg m_e^2$.
The resulting scattering cross section has 
been constrained using the results from XENON10 \cite{Angle:2011th}, XENON100 \cite{Aprile:2016wwo}, DarkSide-50 \cite{Agnes:2018oej} and SENSEI \cite{Crisler:2018gci}. Additionally,  low energy accelerator experiments impose constraints on the $U(1)_{B-L}$ gauge field  mass and coupling \cite{Bauer:2018onh,Ilten:2018crw}. Although they are evaluated for a gauge field with kinetic mixing term, the constraints can be easily translated to the present model. The parameter space is very constrained for the range of masses presented in Table \ref{tab:relicDens}  (40 -- 700 MeV). Only the BM II with $\delta_A=10$ is not ruled out by direct detection experiments. Of course, these parameter choices are representative and other values of $R$, $L$ and $\delta_A$ can give similar results.

In Fig. \ref{fig:plotlimitBM2delta10} we show the part of the parameter space that is allowed to explain the observed DM relic density.

\section{ Conclusions}\label{sec:discussion}

In this paper we have investigated the role of an abelian gauge field as  mediator of the interaction between a DM candidate and the SM, in a model with two ED compactified on the chiral square. DM is localized in one thin brane at the conical singularity $(0,0)$, while a fat brane is lying  between $(\pi r, \pi r)$ and $(\pi R, \pi R)$. SM is confined in the fat brane and its fields propagate in the ED similarly to UED models, but the vector mediator interacts with the visible sector only though  $B-L$ charged particles. 

BLKT is present only in the thin brane because the BC do not allow them in the thick brane \cite{Landim:2019ufg}. Notice that we did not need to introduce a kinetic-mixing term and the relative smallness of the coupling can be explained if a BLKT in the fat brane is very small and can be neglected.  Due to BC the $U(1)$ symmetry is broken   without demanding any Higgs mechanism in the bulk, and the  resulting roots that determine the masses of the KK states depend only upon the BLKT parameter $\delta_A$. The effective coupling between the mediator and the SM particles, due to the fat brane, has a similar behavior as in previous results \cite{Landim:2019epv}, depending upon the 6-D compactification radius $R$ and the SM brane thickness $L$, although it is not as suppressed as in the 5-D case.  Considering a complex scalar field as a DM candidate, the DM relic abundance can be satisfied by some parameter choices, whose values also avoid direct detection constraints.  

This model may lead to distinct signatures in the upcoming experiments and it resembles the 5-D case: the combination of searches for KK vector mediators  and UED  particles in two ED,  where for the latter the compactification radius $L$ is constrained through the missing energy from the cascade 
decay of SM KK particles. Moreover, the  main final states of the lightest mediator decay  are     missing energy or charged  leptons, while for the KK mediator modes, twice or more as heavy as DM,  the resulting cascade decay  gives a missing energy signature as well.

\begin{acknowledgements}
 We thank the anonymous referee for insightful comments. This work was supported by CAPES under the process 88881.162206/2017-01 and Alexander von Humboldt Foundation. \end{acknowledgements}

\bibliographystyle{unsrt}
\bibliography{references}\end{document}